\documentclass[aps,prd,onecolumn,superscriptaddress,nofootinbib,amsmath,amssymb]{revtex4}

\usepackage{graphicx}
\usepackage{wrapfig}

 \tighten  \usepackage[hypertexnames=false]{hyperref}
\begin{document}

 \title{Asymptotically flat spacetimes in 3D bigravity.}

\author{Hern\'an A. Gonz\'alez}
\email{hdgonzal-at-uc.cl}
\affiliation{Departamento de F\'{\i}sica,\\
P. Universidad Cat\'{o}lica de Chile, Casilla 306, Santiago 22,Chile. }

\author{Miguel Pino}
\email{mnpino-at-uc.cl}
\affiliation{Departamento de F\'{\i}sica,\\
P. Universidad Cat\'{o}lica de Chile, Casilla 306, Santiago 22,Chile. }

\begin{abstract}
We report that a class of three-dimensional bimetric theories contain asymptotically flat solutions. These spacetimes can be cast in a set of asymptotic conditions at null infinity which are preserved under the infinite dimensional BMS group. Moreover, the algebra of the canonical generators exhibits a central extension. The possibility that these solutions describe regular black holes is also discussed.   
\end{abstract}

\maketitle

\section{Introduction}

A property of some asymptotic spacetime symmetry groups is that sometimes they are much bigger than the group of isometries of the background geometry. The first example of this behavior was observed in four dimensional asymptotically flat spacetimes. There, the background geometry is Minkowski, and its corresponding isometry group, the Poincar\'e group, gets enhanced to the infinite dimensional BMS group at null infinity \cite{Bondi:1962px,Sachs:1962zza}. 

A particularly remarkable case was found by Brown and Henneaux \cite{Brown:1986nw}. They showed that asymptotic symmetry of Anti-de Sitter geometry in 2+1 dimensions consist of two copies of the Virasoro algebra. Moreover, they found that a classical central extension emerged in the algebra of the canonical generators. This fact was used to give a statistical interpretation to the entropy of the BTZ black hole \cite{Banados:1992wn} using conformal field theory methods \cite{Strominger:1997eq}.  The central charge plays a fundamental role in this analysis. In a sense, it gives a measure of the number of degrees of freedom per spacetime point of the theory.

A similar structure for asymptotically flat spacetimes has been discovered recently, giving some hope that the entropy of this kind of black holes may be also accounted by using these techniques. First, in \cite{Ashtekar:1996cd}, the BMS group of symmetries was defined for asymptotically flat spacetimes in three dimensions. The central extension of its algebra was discovered in \cite{Barnich:2006av}. More recently, in \cite{Barnich:2011mi}, a field dependent central extension of the $bms_4$ algebra was found. For the Kerr black hole it turned out to be proportional to the angular momentum. However, a microscopical description of the entropy of the black hole in asymptotically flat spaces is still missing. The obstacle may be spotted from the situation in three dimensions. Despite its similarities with the asymptotic structure of the AdS case, the latter techniques cannot give rise to horizon thermodynamics simply due to the lack of an asymptotically flat black hole in general relativity. 

Nonetheless, this kind of solutions can be build up from modifications of Einstein gravity. For instance, in \cite{Oliva:2009ip}, an asymptotically flat black hole was found in New Massive Gravity theory \cite{Bergshoeff:2009hq}. In this sense, another interesting modification to GR consist in two interacting spin two fields and is usually called \emph{bigravity}. There are much literature and recent work on this theory (see \cite{ISS} for the original paper). For recent reviews and new results, including black holes properties, cosmological implications and its relation with massive gravity, see \cite{Blas:2007zz,Clifton:2011jh,Banados:2011hk,DamourKogan,ArkaniHamed:2002sp} and references there in and there off.     

Here we want to report that bimetric theories contain a class of asymptotically flat solutions in three dimensions. These spacetimes can be cast in a set of asymptotic conditions at null infinity  which are preserved under the BMS group. Moreover, the algebra of the canonical generators has a central extension. This central charge can be understood from the contraction of two copies of Virasoro algebra which appear in its AdS counterpart \cite{Banados:2011sd}.

The plan of the paper is organized as follows. In section II a brief introduction to bigravity is given. In section III an asymptotically flat solution is displayed along with its asymptotic Killing vectors. In section IV the canonical charges and its algebra is calculated. Section V discusses the possibility that the asymptotic flat solutions describe regular black holes. 

\section{Bigravity action}

Bigravity is a theory with two independent {\it dynamical} metrics, denoted  by $g_{\mu\nu}$ and $f_{\mu\nu}$. The action, first considered in \cite{ISS}, is
\begin{eqnarray}\label{I}
I[g_{\mu\nu},f_{\mu\nu}] &=& {1 \over 16\pi G} \int d^3x \Big[\ \sqrt{-g} R(g) + \sigma\, \sqrt{-f} R(f)- U(g,f) \Big]
\end{eqnarray}
where $R(g)$ and $R(f)$ are the Ricci scalars associated to each metric and $U(g,f)$ is an interaction potential depending both on $g$ and $f$. The dimensionless parameter $\sigma$ measures the relative strengths of both Newton's constants. 

If the potential were absent, the symmetry group would be two copies of diffeomorphisms, one for each metric. However, the potential is constructed in such a way that it breaks down the symmetry to just one diffeomorphism acting simultaneously on both fields.   

For a large class of potentials, flat space $g_{\mu\nu}=f_{\mu\nu} =\eta_{\mu\nu}$ is a solution to the equations of motion. Furthermore, the linearized theory around this background gives a Pauli-Fierz term (see \cite{DamourKogan} for details).

In particular, the potential we will use is \footnote{We used this potential in order to make contact with \cite{Banados:2011hk} and \cite{Banados:2011sd}.}
\begin{eqnarray}\label{U2}
U(g,f) &=& \sqrt{-f}( g_{\mu\nu}-f_{\mu\nu}) (g_{\alpha\beta}-f_{\alpha\beta} )\times \nonumber\\
&& \ \ \ \ \ \   \Big[p^2_1 ( f^{\mu\alpha} f^{\nu\beta} - f^{\mu\nu}f^{\alpha\beta}) -p^2_2( g^{\mu\alpha}g^{\nu\beta} - g^{\mu\nu}g^{\alpha\beta})\Big].
\end{eqnarray}
where $g^{\mu\nu}$ and $f^{\mu\nu}$ are the inverses of $g_{\mu\nu}$ and $f_{\mu\nu}$, respectively. The parameters $p_1$ and $p_2$ have dimension of mass. 

\section{Asymptotically flat solution and fall-off conditions}

A solution for the system \eqref{I} with axial symmetry is
\begin{eqnarray}
  g_{\mu\nu}dx^\mu dx^\nu\equiv&dg^2&= -h(r)dt^2 + {dr^2 \over g(r)} + 2 J dt d\varphi+ r^2 d\varphi^2, \label{metricg} \\
  f_{\mu\nu}dx^\mu dx^\nu\equiv&df^2&= -X(r)dt^2 + Y(r) dr^2 + 2J dt d\varphi+ 2H(r)dt dr +  r^2 d\varphi^2, \label{metricf} 
\end{eqnarray}
where 
\begin{eqnarray}
h(r)&=&M_g+{\sigma C \over r}\\
g(r)&=&M_g+{\sigma C \over r}+{J^2 \over r^2}\\
X(r)&=&M_f-{ C \over r}\\
H(r)^2&=&1-Y(r)\left( X(r)+{J^2 \over r^2}\right)
\end{eqnarray}
and function $Y(r)$ is best expressed by the relation
\begin{equation}
{X(r)+{J^2 \over r^2} \over g(r)}+Y(r) g(r) = 1+\frac{p_1}{p_2}+\frac{\sigma C}{4 p_2^2 r^3},
\end{equation}
where $M_g$, $M_f$, $J$ and $C$ are arbitrary integration constants.

Each metric corresponds to asymptotically locally flat spacetime, i.e., the Riemann tensors $R^{\mu \nu}_{\rho \sigma}(g)$ and $R^{\mu \nu}_{\rho \sigma}(f)$ vanish for large values of the radial coordinate. Indeed, as we will see below, spacetimes \eqref{metricg} and \eqref{metricf} belong to a more general set of solutions that approach Minkowski space in the region $r\rightarrow \infty$. 

From the conformal point of view, the right boundary to impose asymptotic conditions is null infinity \cite{Witten:2001kn}. In turn, the latter solutions can be accommodated in a set of fall-off conditions at future null infinity preserved under the action of the isometries of flat space. In doing so, we will use the change of coordinates
\begin{eqnarray}\label{dt}
dt&=&du+\frac{dr}{h(r)+\frac{J^2}{r^2}},\label{dt}\\
d\varphi&=&d\phi-\frac{J}{h(r)r^2+J^2}dr.\label{dphi}
\end{eqnarray} 
In this new set of coordinates $u$, $r$ and $\phi$, both fields can be written as perturbations over the background $\bar{g}_{\mu \nu}dx^{\mu}dx^{\nu}=-du^2-2dudr+r^2d\phi^2$, i.e., $g_{\mu \nu}=\bar{g}_{\mu \nu}+\Delta g_{\mu \nu}$ and $f_{\mu \nu}=\bar{g}_{\mu \nu}+\Delta f_{\mu \nu}$. The background $\bar{g}_{\mu\nu}$ corresponds to Minkowski space foliated by outgoing null rays. 

The appropriated fall-off conditions are
\begin{subequations}\label{sols}
\begin{align}
\Delta g_{u u}&=h_{u u}^{g}(u,\phi)+\cdots&\Delta f_{u u}&=h_{u u}^{f}(u,\phi)+\cdots\\
\Delta g_{u r}&=\frac{h_{u r}^{g}(u,\phi)}{r^2}+\cdots&\Delta f_{u r}&=\frac{h_{u r}^{f}(u,\phi)}{r^2}+\cdots\\
\Delta g_{u \phi}&=h_{u \phi}^{g}(u,\phi)+\cdots&\Delta f_{u \phi}&=h_{u \phi}^{f}(u,\phi)+\cdots\\
\Delta g_{r r}&=\frac{h_{r r}^{g}(u,\phi)}{r^3}+\cdots&\Delta f_{r r}&=\frac{h_{r r}^{f}(u,\phi)}{r^3}+\cdots\\
\Delta g_{r \phi}&=\frac{h_{r \phi}^{g}(u,\phi)}{r^2}+\cdots&\Delta f_{r \phi}&=\frac{h_{r \phi}^{f}(u,\phi)}{r^2}+\cdots\\
\Delta g_{\phi \phi}&=h_{\phi \phi}^{g}(u,\phi)+\cdots&\Delta f_{\phi \phi}&=h_{\phi \phi}^{f}(u,\phi)+\cdots
\label{fall}
\end{align}
\end{subequations}

In order for $f_{\mu\nu}$ \eqref{metricf} fulfills these conditions, we have restricted the parameters appearing in the potential $U(g,f)$ to $p_1=p_2$.

It can be seen that the asymptotic behavior \eqref{sols} is invariant under a coordinate transformations generated by the following vectors
\begin{subequations}\label{vect}
\begin{eqnarray}
\xi^u&=&T(\phi)+u\partial_\phi L(\phi)+\cdots\\
\xi^r&=&-r\partial_\phi L(\phi)+\partial_\phi^2 T(\phi)+u \partial_\phi^3 L(\phi)+\cdots\\
\xi^\phi&=&L(\phi)-\frac{1}{r}\left( \partial_\phi T(\phi)+u \partial_\phi^2 L(\phi)  \right)+\cdots
\end{eqnarray}
\end{subequations}
where $T(\phi)$ and $L(\phi)$ are arbitrary functions of $\phi$ and the ellipsis stands for terms which are sub-leading in \eqref{vect}. These vectors form a closed algebra under Lie brackets, in the sense of $[\xi_1,\xi_2]=\xi_{[1,2]}$. The arbitrary functions representing $\xi_{[1,2]}$ reads
\begin{eqnarray}
L_{[1,2]}&=&L_1\partial_{\phi}L_2-L_2\partial_{\phi}L_1\label{L},\\
T_{[1,2]}&=&T_1\partial_{\phi}L_2+L_1\partial_{\phi}T_2-T_2\partial_{\phi}L_1-L_2\partial_{\phi}T_1.\label{T}
\end{eqnarray}
By defining Fourier modes $t_m=\xi(T=e^{in\phi},L=0)$ and $j_m=\xi(T=0,L=e^{in\phi})$ the algebra looks
\begin{eqnarray}
i[t_m,t_n]=0, \quad i[j_m,t_n]=(m-n)t_{m+n}, \quad i[j_m,j_n]=(m-n)j_{m+n}.
\end{eqnarray}
This is a Lie algebra which consists of a semi-direct sum of diffeomorphisms on the circle generated by $j_m$ and supertranslations $t_m$. It is known as $bms_3$. Note that the modes $m=0,1,-1$ represent the exact Killing vectors of Minkowski space and form a subalgebra which correspond to the Poincar\'e algebra in three dimensions. 

In sum, vectors \eqref{vect} form a representation of the $bms_3$ Lie algebra under Lie brackets. 

\section{Canonical charges and central extension}

Using the Regge-Teitelboim approach \cite{ReggeTeitelboim}, we compute the variation of the charge associated to the asymptotic conditions \eqref{sols} and \eqref{vect}, which turns out to be integrable and gives
\begin{equation}\label{charge}
Q[T,L]=\frac{1}{16\pi G}\int_0^{2\pi} d\phi\left\{\left( T+u\partial_\phi L\right)\big[ h_{uu}^g+\sigma h_{uu}^f\big]+2L\big[h_{u \phi}^g+\sigma h_{u \phi}^f\big]\right\},
\end{equation}
where we have set $Q$ to give zero when evaluated in the background.

It is worth to make a few comments on the calculation of \eqref{charge}. Since outgoing null coordinates are not suitable for Hamiltonian ADM formulation, we have modified the coordinates introducing a regulator $\epsilon$ in such a way that Minkowski space becomes $ds^2=-du^2+(1-\epsilon^2)dr^2-2\epsilon du dr+r^2d\phi^2$. As $\epsilon \rightarrow 1$ we recover outgoing null coordinates. We compute the charges using these coordinates, taking the limit at the end. Although we do not have a proof that this method is strictly correct, it can be used to calculate the mass of known solutions in GR, such Schwarzschild black holes, giving the right value. Furthermore, in \cite{Barnich:2001jy} an independent method to compute the charges was given. The regularization procedure used here and the method in \cite{Barnich:2001jy} gives the same result.   

It is important to notice that the charge $Q$ is conserved only when the following relations hold
\begin{equation}
\partial_u h_{uu}^g=0, \quad \partial_\phi h^g_{uu}-2\partial_uh^g_{u\phi}=0,
\end{equation}
and the same relation are valid for $h^f_{uu}$ and $h^f_{u\phi}$. These equations are the leading order of the $[u,u]$ and $[u,\phi]$ components of the asymptotic equations of motions. 
 
Now we turn to compute the algebra of these canonical charges. The Dirac brackets $\{Q[\xi_1],Q[\xi_2]\}$ can be interpreted as the variation of $Q[\xi_1]$ along a deformation given by $\xi_2$. So, by using \eqref{charge}, we evaluate the algebra of the charges
\begin{equation}\label{alg1}
\delta_{\xi_2} Q[\xi_1] = Q[\xi_{[1,2]}]+K[\xi_1,\xi_2],
\end{equation} 
$K[\xi_1,\xi_2]$ is a central extension. By straightforward calculation, it gives
\begin{equation}
K[\xi_1,\xi_2]=\frac{1+\sigma}{8\pi G}\int_0^{2\pi}d\phi \left\{ \partial_\phi L_1 \left( \partial_\phi^2 T_2+T_2\right)- \partial_\phi L_2 \left( \partial_\phi^2 T_1+T_1\right) \right\}. 
\end{equation}
It is instructive to write algebra \eqref{alg1} in Fourier modes by using the definitions $J_{m}=Q[0,j_m]$ and $T_{m}=Q[0,t_m]$
\begin{subequations}\label{alg2}
\begin{eqnarray}
i\{T_{m},T_{n}\}&=&0,\\
i\{J_{m},T_{n}\}&=&(m-n)T_{n+m}+\frac{1+\sigma}{4G}n(n^2-1)\delta_{n+m},\label{central}\\
i\{J_{m},J_{n}\}&=&(m-n)J_{n+m}.
\end{eqnarray}
\end{subequations}
and we can see that corresponds to $bms_3$ with a central extension. It is straightforward to check that this algebra fulfills Jacobi identity.
 
Two comment can be made. First, it is suggestive to write the central charge in \eqref{central} as $$\frac{3}{G}+\frac{3}{G/\sigma}.$$  As it was realized in \cite{Barnich:2006av} for three dimensional General Relativity, an asymptotically flat spacetime at null infinity picks up a central charge given by ${3 \over G}$ . Hence, for the present theory, the central charge obtained in \eqref{central} can be regarded as having two contributions, one from $g$ and one from $f$. This can be understood from the fact that there is no derivative terms in the potential \eqref{U2}. This observation was already made in \cite{Banados:2009it} for asymptotically AdS black holes in bigravity.     

Another interesting comment is that \eqref{alg2} can be understood as a contraction of two Virasoro algebras
\begin{subequations}\label{alg3}
\begin{eqnarray}
i\{L^{\pm}_{m},L^{\pm}_{n}\}&=&(m-n)L^{\pm}_{n+m}+\frac{\ell(1+\sigma)}{8G}n(n^2-1)\delta_{n+m},\\
i\{L^{\pm}_{m},L^{\mp}_{n}\}&=&0.
\end{eqnarray}
\end{subequations}
It is easy to show, that by defining $T_{m}=\ell^{-1}\left(L^{+}_{m}+L^{-}_{-m}\right)$ together with $J_{m}=L^{+}_{m}-L^{-}_{-m}$ and taking $\ell \rightarrow \infty$, we recover \eqref{alg2}.
Actually, a realization of \eqref{alg3} was found in the charge algebra of asymptotically AdS solutions for a bigravity theory with the same potential considered here, \cite{Banados:2011sd}. 

\section{Black holes?}
In this section we analyze the geometry generated by \eqref{metricg} and \eqref{metricf} and study the conditions needed to describe regular black holes. When considered independently, is easy to see that $g_{\mu\nu}$ has an horizon where $g(r_g)=0$,  and $f_{\mu\nu}$ has one in $X(r_f)+\frac{J^2}{r_f^2}=0$. In order to describe black holes, both functions $g(r)$ and $X(r)+\frac{J^2}{r^2}$ must be positive for large values of $r$. This is achieved by imposing
\begin{equation}\label{bhconditions}
M_g >0 \quad M_f>0 \quad Q>0 \quad -1<\sigma<0
\end{equation} 

In the coordinates \eqref{dt} and \eqref{dphi}, line elements $df^2$ and $dg^2$ posses two Killing vectors $\partial_{u}$ and $\partial_{\phi}$. They can be written in the form \eqref{vect}. Using \eqref{charge}, their respective conserved charges are
\begin{equation}
\mathcal{E}=\frac{1}{8G}\left[\left(1-M_g\right)+\sigma \left(1-M_f\right)\right], \quad \mathcal{J}=\frac{(1+\sigma)}{4G}J
\end{equation}
We identify $\mathcal{E}$ with the gravitational energy of this solution and $\mathcal{J}$ with its angular momentum. (In the sense of \cite{ReggeTeitelboim}, coordinate transformation \eqref{dt} \eqref{dphi} is a proper gauge transformation, i.e., it does not affect the conserved charges).

It is straightforward to check that the positivity of the energy $\mathcal{E}>0$ is compatible with conditions \eqref{bhconditions} for some range of $M_g$ and $M_f$.

A more interesting condition appears if we study the near-horizon geometry of this solution. It has been shown in \cite{Banados:2011hk,Deffayet:2011rh} that for any bigravity theory, in order to have well behaved fields in all points of spacetime, both event horizons, $r_g$ and $r_f$, must coincide and the surface gravities of $g_{\mu\nu}$ and $f_{\mu\nu}$ must be the same. The former condition is easily fulfilled by adjusting $M_g$ and $M_f$ as
\begin{equation}
M_g=\frac{|\sigma|C}{r_0}-\frac{J^2}{r_0^2} \quad\quad M_f=\frac{C}{r_0}-\frac{J^2}{r_0^2}
\end{equation}
where $r_0$ is the location of the horizon. 

The latter condition can be achieved in two different ways:
\begin{itemize}
\item $C=0$. Both fields becomes equal and no longer have horizons. The outcoming spacetimes have causal structures reminiscent of cosmologies. 
\item $\sigma=-1$. Notice that with this condition $M_g=M_f$ and both fields describe the same black hole written in different coordinates, i.e., with an appropriated change of coordinates, $f$  acquires the form of $g$ and viceversa. Besides, all global charges vanish, in particular, its entropy (see \cite{Banados:2011hk} for a proper definition of entropy in bigravity). This value for the entropy may have relation with the fact that for $\sigma=-1$ the value of the central charge \eqref{central} is zero.
\end{itemize} 

\section{Conclusions}

A class of solution to a bigravity theory has been presented. These solutions can be accommodated in a set of asymptotic conditions at null infinity which are preserved under the action of the BMS group, meaning that these solutions are asymptotically flat. However, in order to satisfy this latter condition the parameters appearing in the potential $U(f,g)$ have to fulfill $p_1=p_2$.

The algebra of the canonical generators was calculated and a central extension can be found giving $\frac{3}{G}\left(1+\sigma\right)$. This algebra was understood from the contraction of two Virasoro algebras for asymptotically AdS solutions for the same theory.

It was shown that, due to regularity conditions, the solution which describes an asymptotically flat and regular black hole is produced only when $\sigma=-1$. However, this condition sets the central extension to zero. It is tempting to relate the latter with the vanishing total entropy of this solution.

As a remark, this result does not exclude the possibility that there exists regular solutions to bigravity with the same asymptotic behavior and a non vanishing central extension. Perhaps this can be achieved by using other potentials. Building solutions satisfying these features is a problem that will be considered elsewhere.

Finally, it would be desirable to study the asymptotic symmetries of bigravity solutions in four dimensions and to see if it is possible to find a $bms_4$ charge algebra, as it was done in \cite{Barnich:2010eb,Barnich:2009se}.

\section{Acknowledgments}
The authors would like to thank M\'aximo Ba\~nados, Glenn Barnich and David Tempo for useful discussions. HG thanks Conicyt for financial support.


\begin{thebibliography}{10}

\bibitem{Bondi:1962px}
  H.~Bondi, M.~G.~J.~van der Burg and A.~W.~K.~Metzner,
  Proc.\ Roy.\ Soc.\ Lond.\ A {\bf 269} (1962) 21.
  
\bibitem{Sachs:1962zza}
  R.~Sachs,
  Phys.\ Rev.\  {\bf 128} (1962) 2851.
  
\bibitem{Brown:1986nw} 
  J.~D.~Brown and M.~Henneaux,
  Commun.\ Math.\ Phys.\  {\bf 104}, 207 (1986).
  
\bibitem{Banados:1992wn} 
  M.~Banados, C.~Teitelboim and J.~Zanelli,
  Phys.\ Rev.\ Lett.\  {\bf 69}, 1849 (1992)
  [hep-th/9204099].
  
\bibitem{Strominger:1997eq} 
  A.~Strominger,
  JHEP {\bf 9802}, 009 (1998)
  [hep-th/9712251].
  
\bibitem{Ashtekar:1996cd} 
  A.~Ashtekar, J.~Bicak and B.~G.~Schmidt,
  Phys.\ Rev.\ D {\bf 55}, 669 (1997)
  [gr-qc/9608042].
  
\bibitem{Barnich:2006av} 
  G.~Barnich and G.~Compere,
  Class.\ Quant.\ Grav.\  {\bf 24}, F15 (2007)
  [gr-qc/0610130].
  
\bibitem{Barnich:2011mi} 
  G.~Barnich and C.~Troessaert,
  JHEP {\bf 1112}, 105 (2011)
  [arXiv:1106.0213 [hep-th]].

\bibitem{Oliva:2009ip} 
  J.~Oliva, D.~Tempo and R.~Troncoso,
  JHEP {\bf 0907}, 011 (2009)
  [arXiv:0905.1545 [hep-th]].

\bibitem{Bergshoeff:2009hq} 
  E.~A.~Bergshoeff, O.~Hohm and P.~K.~Townsend,
  Phys.\ Rev.\ Lett.\  {\bf 102}, 201301 (2009)
  [arXiv:0901.1766 [hep-th]].

\bibitem{ISS}
C.~J. Isham, Abdus Salam, and J.~A. Strathdee.
\newblock {F-dominance of gravity}.
\newblock {\em Phys. Rev.}, D3:867--873, 1971.

\bibitem{Blas:2007zz} 
  D.~Blas, C.~Deffayet and J.~Garriga,
  Phys.\ Rev.\ D {\bf 76}, 104036 (2007)
  [arXiv:0705.1982 [hep-th]].

\bibitem{Clifton:2011jh} 
  T.~Clifton, P.~G.~Ferreira, A.~Padilla and C.~Skordis,
  Phys.\ Rept.\  {\bf 513}, 1 (2012)
  [arXiv:1106.2476 [astro-ph.CO]].

\bibitem{Banados:2011hk} 
  M.~Banados, A.~Gomberoff and M.~Pino,
  Phys.\ Rev.\ D {\bf 84}, 104028 (2011)
  [arXiv:1105.1172 [gr-qc]].

\bibitem{DamourKogan}
Thibault Damour and Ian~I. Kogan.
\newblock {Effective Lagrangians and universality classes of nonlinear
  bigravity}.
\newblock {\em Phys. Rev.}, D66:104024, 2002.

\bibitem{ArkaniHamed:2002sp} 
  N.~Arkani-Hamed, H.~Georgi and M.~D.~Schwartz,
  Annals Phys.\  {\bf 305}, 96 (2003)
  [hep-th/0210184].

\bibitem{Banados:2011sd} 
  M.~Banados and M.~Pino,
  arXiv:1112.0042 [gr-qc].

\bibitem{Witten:2001kn} 
  E.~Witten,
  hep-th/0106109.

\bibitem{ReggeTeitelboim}
Tullio Regge and Claudio Teitelboim.
\newblock {Role of Surface Integrals in the Hamiltonian Formulation of General
  Relativity}.
\newblock {\em Ann. Phys.}, 88:286, 1974.

\bibitem{Barnich:2001jy} 
  G.~Barnich and F.~Brandt,
  Nucl.\ Phys.\ B {\bf 633}, 3 (2002)
  [hep-th/0111246].

\bibitem{Banados:2009it} 
  M.~Banados and S.~Theisen,
  JHEP {\bf 0911}, 033 (2009)
  [arXiv:0909.1163 [hep-th]].

\bibitem{Deffayet:2011rh} 
  C.~Deffayet and T.~Jacobson,
  Class.\ Quant.\ Grav.\  {\bf 29}, 065009 (2012)
  [arXiv:1107.4978 [gr-qc]].

\bibitem{Barnich:2010eb} 
  G.~Barnich and C.~Troessaert,
  JHEP {\bf 1005}, 062 (2010)
  [arXiv:1001.1541 [hep-th]].
   
\bibitem{Barnich:2009se}
  G.~Barnich and C.~Troessaert,
  Phys.\ Rev.\ Lett.\  {\bf 105} (2010) 111103
  [arXiv:0909.2617 [gr-qc]].
  
\end{thebibliography}
\end{document}